# Solar Sail Propulsion by 2050:
# An Enabling Capability for Heliophysics Missions

Les Johnson[1], Nathan Barnes[2], Matteo Ceriotti[3], Thomas Y. Chen[4], Artur Davoyan[5], Louis Friedman[6], Darren Garber[7], Roman Kezerashvili[8], Ken Kobayashi[9], Greg Matloff[8], Colin McInnes[10], Pat Mulligan[11], Grover Swartzlander[12], Slava G. Turyshev[13]

[1] NASA George C. Marshall Space Flight Center, [2] L'Garde, Inc., [3] University of Glasgow, [4] Columbia University, [5] University of California, Los Angeles, [6] The Planetary Society, [7] Nextrac, Inc. [8] New York City College of Technology, [9] NASA George C. Marshall Space Flight Center, [10] University of Glasgow, [11] NOAA, [12] Rochester Institute of Technology, [13] NASA Jet Propulsion Laboratory

**INTRODUCTION, BENEFITS, and STATE-OF-THE-ART**

Solar sails provide extremely large ΔV (potentially many 10s of km/sec) enabling new vantage points for Heliophysics observations that are inaccessible or not practically implementable using conventional rocket propulsion. Solar sails obtain thrust by using lightweight materials that reflect sunlight to propel a spacecraft. The continuous photon pressure from the Sun's rays provides thrust, with no need for the heavy, expendable primary propellants employed by conventional on-board chemical and electric propulsion systems that limit mission lifetime and observation locations. Sails are relatively low-cost to implement and the continuous solar photon pressure provides propellantless thrust to perform a wide range of advanced maneuvers, such as to hover indefinitely at points in space, or conduct high ΔV orbital plane changes. Solar sail propulsion systems can also propel a space vehicle to tremendous speeds—theoretically much faster than any present-day propulsion system. With solar propulsive energy, sails are propellantless, thereby significantly increasing useful payload mass.

Solar sails enable missions to observe the solar environment from unique vantage points, such as sustained observations away from the Sun-Earth line (SEL); sustained sub-L1 (sunward of L1 along the SEL) station keeping for improving space-weather monitoring, prediction, and science,[1] advancing a capability within the range of interest to NOAA,[2] and supporting human spaceflight crew safety and health needs[3]; sustained in-situ Earth magnetotail measurements[4]; and, in the mid-term, those that require a high inclination solar orbit[5,6]; Earth polar-sitting and polar viewing observatories;[7] low perihelion and out-of-the-ecliptic missions; multiple fast transit missions to study heliosphere to interstellar medium transition, as well as missions of interest across a broad user community.[8] Additionally, sail trajectory observation may provide a test of the fundamental principles of general relativity and a special relativistic effect, known as the Poynting–Robertson effect.[16a] The curvature of spacetime, in conjunction with solar radiation pressure, affects bound orbital sail motion and leads to deviations from Kepler's third law for heliocentric and non Keplerian orbits[16b], as well as to the new phenomenon for non-Keplerian orbits when the orbital plane precesses around the sun that is an analog of the Lense-Thirring effect.[16b,9]

In the early-to-mid 2000's, NASA's In-Space Propulsion Technology Project developed and tested two different 400 $m^2$ solar sail systems in the Glenn Research Center's Space Power Facility at Plum Brook Station, Ohio. In the summer of 2010, the Japanese Aerospace Exploration Agency, JAXA, launched a solar sail spacecraft named IKAROS in tandem with another mission to Venus. The IKAROS was the first in-flight demonstration of solar sailing.[10] Numerous program objectives were achieved and included verification of solar radiation pressure effects and validation of in-flight guidance and navigation techniques.

The goals of the NanoSail-D mission were demonstration of both the deployment of a 10-$m^2$ sail from a three-unit CubeSat and the use of sail-induced acceleration for reducing de-orbit time. NASA's Marshall Space Flight Center (MSFC) and Ames Research Center (ARC) collaborated with MSFC responsible for the solar sail assembly and ARC responsible for the bus. NanoSail-D was launched in November 2010 and deployed in January 2011. The uncontrolled NanoSail-D spacecraft was expected to deorbit from the initial 650 km orbit altitude in 70–120 days, but it remained in orbit for 240 days prior to reentry.[11,12]

The LightSail 1 and 2 missions were the culmination of a decade-long program sponsored by the privately-funded Planetary Society. The LightSail 1 was launched into a 55°, 356 km × 705 km elliptical orbit where the 32 $m^2$ sail was successfully deployed. LightSail 2, launched in 2019, demonstrated controlled solar sailing in LEO performing 90 degree slews in each orbit to harness momentum from solar photons. Flight data show that LightSail 2 successfully controlled its orientation relative to the Sun, and the controlled thrust from solar radiation pressure measurably reduced the rate of orbital decay.[13]

The NASA Near Earth Asteroid Scout (NEAS) mission is led by NASA MSFC with support from JPL. Planned for launch on the Space Launch System Artemis 1 mission in 2021, NEAS will use an 86-$m^2$ square solar sail to propel the 6U CubeSat bus on a reconnaissance flyby trajectory of a 100 m asteroid. NEAS represents the first space science mission that will utilize solar sailing to achieve its science objectives.[14] The LaRC Advanced Composite Solar Sail System (ACS3) will demonstrate deployment of an approximately 74-$m^2$ composite boom solar sail system in low-Earth orbit after 2021. The MSFC Solar Cruiser mission (now in competitive Phase A) will mature solar sail technology for use in future Heliophysics missions during its 2024 flight. If selected, Solar Cruiser mission will fly a >1600 $m^2$ solar sail containing embedded reflectivity control devices and photovoltaic cells. The mission timeline includes deployment of largest sail ever flown, validation of all sail subsystems, controlled station-keeping inside of the Sun-Earth L1 point, demonstration of pointing performance for science imaging, and an increase in heliocentric inclination (out of the ecliptic plane).[15]

A mission to provide high-resolution, multi-pixel images of exoplanets has been studied for several years in a series of increasing NIAC studies (now in phase III) to reach and use the Solar Gravity Lens, beyond 600 AU from Earth. The mission would achieve escape velocities from the solar system > 100 km/s. An inner solar system Technology Demonstration Mission with a Solar Sail and Smallsat is proposed as part of this study.[16][17][18]

**CURRENT TECHNOLOGY AND NEAR-FUTURE IMPROVEMENTS:**

NanoSail-D, the LightSails, and development of NEAS and ACS3 provide the technology base for practical implementation of 3-12U CubeSat class missions. With the use of composite booms, NEAS-class sails with areas of 200 – 300 $m^2$ are now realizable. The next step is the demonstration of the technologies needed to fabricate, fly, navigate and control 100 – 200 kg class missions using sails in the 1,500 to 7,000-$m^2$ class. These are being developed and ground tested and are ready for demonstration in the proposed Solar Cruiser[19].

**MID-TERM TECHNOLOGIES FOR MORE AMBITIOUS MISSIONS:**

Close solar approach in future missions will require novel solar sail materials. New Liquid Crystal Polymer film technology appears to possess the very low absorption manufacturability and toughness required. The long-chain molecules in these nematic crystals are optically birefringent, slowing light polarized along the long axis of the molecule compared to the short axis, their properties can be tailored to affect the force, and torque response without a change is macroscopic shape with lower power requirements than other methods.[20][21]

Monolayer graphene and graphene based composites offer advantages for these future systems. Graphene, for example, has low aerial density and provides much higher tensile strength[22] and temperature capability up-to 4,000 K. Atomically thin graphene is examined as a promising stand-alone material for solar sails owing to its extremely low areal density making it ideal for extremely large sails that require close perihelion approaches. Other light, high-temperature materials such as aerographite[23] are also under consideration. Further, ceramics, such as silicon nitride and silicon dioxide, are naturally transparent with very low losses in the visible and ultraviolet. These materials are refractory and possess high melting points (>2000K) making them potentially applicable for close perihelion approaches (4 − 20 solar radii) with minimal degradation. Close approach can be used to accelerate solar sails to unprecedented velocities - for example, a sail with a characteristic acceleration $\simeq 3$ mm/s$^2$ performing an Oberth maneuver at 0.1AU perihelion may be accelerated to >20AU/yr cruise velocity and accelerations to 50AU/year are projected with new advance materials.

**CONCLUSION**

Solar sail technology is maturing rapidly and with the recent and planned in-space demonstrations will soon be ready for near-term science mission implementation. By 2050, it is feasible for sail-propelled missions to be studying the sun at many solar latitudes, including polar, sub-L1, and at various locations off the SEL. The first rapid sail-propelled missions to extrasolar destinations will be well underway, soon to return increased knowledge of the interstellar medium and perhaps our first images of exoplanets with magnetospheres. The time to begin planning missions using the enabling propulsive capabilities of solar sails is now.